\documentclass[conference, 10pt, final]{IEEEtran}
\usepackage[nolist]{acronym} 
\usepackage{tikz}
\usepackage{pgfplots}
\usepgfplotslibrary{groupplots}
\usepackage{graphicx}
\usepackage{subfloat}
\usepackage{caption}
\usepackage{subcaption}
\usepackage{float}
\pgfplotsset{compat=newest}
\usepackage{units}
\usepackage{amsmath, amsbsy, amssymb}
\usepackage{tikzscale}
\usetikzlibrary{arrows}
\usepackage{color}
\usepackage{epigraph}
\usepackage{circuitikz}
\usepackage{multirow}
\usepackage{booktabs}
\usepackage[plainruled]{algorithm2e}
\usepackage[group-separator={,}]{siunitx}
\definecolor{darkgreen}{rgb}{0.125,0.5,0.169}
\usetikzlibrary{shapes,arrows}
\usepackage{marvosym}
\usepackage{bbm}
\usepackage{mathtools}

\tikzset{>=latex}

\renewcommand{\vec}[1]{\mathbf{#1}}

\newcommand{\cv}{\vec{c}}

\newcommand{\lv}{\vec{l}}

\newcommand{\uv}{\vec{u}}

\newcommand{\xv}{\vec{x}}
\newcommand{\yv}{\vec{y}}
\newcommand{\zv}{\vec{z}}

\newcommand{\Nc}{{\cal N}}

\newcommand{\RR}{\mathbb{R}}

\newcommand{\LB}{\left(}
\newcommand{\RB}{\right)}
\newcommand{\LP}{\left\{}
\newcommand{\RP}{\right\}}
\newcommand{\LSB}{\left[}
\newcommand{\RSB}{\right]}

\newcommand{\EE}{{\mathbb{E}}}
\newcommand{\Expect}[2]{\EE_{#1}\LSB #2\RSB}

\begin{acronym}
 \acro{CSI}{channel state information}
 \acro{UE}{user equipment}
 \acro{UL}{uplink}
 \acro{BS}{basestation}
 \acro{TDD}{time division duplex}
 \acro{FDD}{frequency division duplex}
 \acro{ECC}{error-correcting code}
 \acro{MLD}{maximum likelihood decoding}
 \acro{HDD}{hard decision decoding}
 \acro{IF}{intermediate frequency}
 \acro{RF}{radio frequency}
 \acro{SDD}{soft decision decoding}
 \acro{NND}{neural network decoding}
 \acro{CNN}{convolutional neural network}
 \acro{ML}{maximum likelihood}
 \acro{GPU}{graphical processing unit}
 \acro{BP}{belief propagation}
 \acro{LTE}{Long Term Evolution}
 \acro{BER}{bit error rate}
 \acro{SNR}{signal-to-noise-ratio}
 \acro{ReLU}{rectified linear unit}
 \acro{BPSK}{binary phase shift keying}
 \acro{QPSK}{quadrature phase shift keying}
 \acro{AWGN}{additive white Gaussian noise}
 \acro{MSE}{mean squared error}
 \acro{LLR}{log-likelihood ratio}
 \acro{MAP}{maximum a posteriori}
 \acro{NVE}{normalized validation error}
 \acro{BCE}{binary cross-entropy}
 \acro{CE}{cross-entropy}
 \acro{BLER}{block error rate}
 \acro{SQR}{signal-to-quantisation-noise-ratio}
 \acro{MIMO}{multiple-input multiple-output}
 \acro{OFDM}{orthogonal frequency division multiplex}
 \acro{RF}{radio frequency}
 \acro{LOS}{line of sight}
 \acro{NLoS}{non-line of sight}
 \acro{NMSE}{normalized mean squared error}
 \acro{CFO}{carrier frequency offset}
 \acro{SFO}{sampling frequency offset}
 \acro{IPS}{indoor positioning system}
 \acro{TRIPS}{time-reversal IPS}
 \acro{RSSI}{received signal strength indicator}
 \acro{MIMO}{multiple-input multiple-output}
 \acro{ENoB}{effective number of bits}
 \acro{AGC}{automated gain control}
 \acro{ADC}{analog to digital converter}
 \acro{ADCs}{analog to digital converters}
 \acro{FB}{front bandpass}
 \acro{FPGA}{field programmable gate array}
 \acro{JSDM}{Joint Spatial Division and Multiplexing}
 \acro{NN}{neural network}
 \acro{IF}{intermediate frequency}
 \acro{LoS}{line-of-sight}
 \acro{NLoS}{non-line-of-sight}
 \acro{DSP}{digital signal processing}
 \acro{AFE}{analog front end}
 \acro{SQNR}{signal-to-quantisation-noise-ratio}
 \acro{SINR}{signal-to-interference-noise-ratio}
 \acro{ENoB}{effective number of bits}
 \acro{AGC}{automated gain control}
 \acro{PCB}{printed circuit board}
 \acro{EVM}{error vector mangnitude}
 \acro{CDF}{cumulative distribution function}
 \acro{MRC}{maximum ratio combining}
 \acro{MRP}{maximum ratio precoding}
 \acro{MRT}{maximum ratio transmission}
 \acro{DeepL}{deep-learning}
 \acro{DL}{deep learning}
 \acro{SISO}{single-input single-output}
 \acro{SGD}{stochastic gradient descent}
 \acro{CP}{cyclic prefix}
 \acro{MISO}{Multiple Input Single Output}
 \acro{LMMSE}{linear minimum mean square error}
 \acro{ZF}{zero forcing}
 \acro{USRP}{universal software radio peripheral}
 \acro{RNN}{recurrent neural network}
 \acro{GRU}{gated recurrent unit}
 \acro{LSTM}{long short-term memory}
 \acro{NTM}{neural turing machine}
 \acro{DNC}{differentiable neural computer}
 \acro{TCN}{temporal convolutional network}
 \acro{FCL}{fully connected layer}
 \acro{MANN}{memory augmented neural network}
 \acro{RNN}{recurrent neural network}
 \acro{DNN}{dense neural network}
 \acro{CNN}{convolutional neural network}
 \acro{FIR}{finite impulse response}
 \acro{BPTT}{back-propagation through time}
 \acro{GAN}{generative adversarial network}
 \acro{ELU}{exponential linear unit}
 \acro{tanh}{hyperbolic tangent}
 \acro{BICM}{bit-interleaved coded modulation}
 \acro{OTA}{over-the-air}
 \acro{IM}{intensity modulation}
 \acro{DD}{direct detection}
 \acro{RL}{reinforcement learning}
 \acro{SDR}{software-defined radio}
 \acro{WGAN}{Wasserstein generative adversarial network}
 \acro{BMD}{bit-metric decoding}
 \acro{BMI}{bit-wise mutual information}
 \acro{LDPC}{low-density parity-check}
 \acro{IDD}{iterative demapping and decoding}
 \acro{JSD}{Jensen-Shannon divergence}
 \acro{MMSE}{minimum mean square error}
 \acro{FFT}{fast Fourier transform}
 \acro{IFFT}{inverse fast Fourier transform}
 \acro{QAM}{quadrature amplitude modulation}
 \acro{EMD}{earth mover's distance}
 \acro{TDL}{tapped delay line}
 \acro{KL}{Kullback-Leibler}
 \acro{PRACH}{physical random access channel}
 \acro{URLLC}{ultra-reliable low-latency communication}
 \acro{ANOMA}{asynchronous non-orthogonal multiple access}
 \acro{FEC}{forward error correction}
 \acro{STE}{straight-through estimator}
 \acro{CRC}{cyclic redundancy check}
\end{acronym}

\definecolor{mittelblau}{RGB}{0, 126, 198}
\definecolor{violettblau}{cmyk}{0.9, 0.6, 0, 0}
\definecolor{rot}{RGB}{238, 28 35}
\definecolor{apfelgruen}{RGB}{140, 198, 62}
\definecolor{gelb}{RGB}{1, 221, 0}
\definecolor{orange}{RGB}{244, 111, 33}
\definecolor{pink}{RGB}{237, 0, 140}
\definecolor{lila}{RGB}{128, 10, 145}
\definecolor{hellgrau}{RGB}{224, 224, 224}
\definecolor{mittelgrau}{RGB}{128, 128, 128}
\definecolor{dunkelgrau}{RGB}{80,80,80}
\definecolor{anthrazit}{RGB}{19, 31, 31}

\pgfplotscreateplotcyclelist{corporate colours markers}{%
rot, every mark/.append style={fill=.!80!rot},mark=*\\%
mittelblau, every mark/.append style={fill=.!80!mittelblau},mark=square*\\%
apfelgruen, every mark/.append style={fill=.!80!apfelgruen},mark=triangle*\\%
orange, mark=star\\%
pink, every mark/.append style={fill=.!80!pink},mark=diamond*\\%
violettblau, every mark/.append style={fill=.!80!violettblau},mark=otimes*\\%
lila, mark=|\\%
gelb, every mark/.append style={fill=.!80!gelb},mark=pentagon*\\%
hellgrau, mark=text,text mark=p\\%
anthrazit, mark=text,text mark=a\\%
}

\pgfplotscreateplotcyclelist{corporate colours markers double}{%
rot, every mark/.append style={fill=.!80!rot},mark=*\\%
rot, dashed, every mark/.append style={fill=.!80!rot,solid},mark=*\\%
mittelblau, every mark/.append style={fill=.!80!mittelblau},mark=square*\\%
mittelblau, dashed, every mark/.append style={fill=.!80!mittelblau,solid},mark=square*\\%
apfelgruen, every mark/.append style={fill=.!80!apfelgruen},mark=triangle*\\%
apfelgruen, dashed, every mark/.append style={fill=.!80!apfelgruen,solid},mark=triangle*\\%
orange, mark=star\\%
orange, dashed, mark=star\\%
pink, every mark/.append style={fill=.!80!pink},mark=diamond*\\%
pink, dashed, every mark/.append style={fill=.!80!pink,solid},mark=diamond*\\%
violettblau, every mark/.append style={fill=.!80!violettblau},mark=otimes*\\%
violettblau, dashed, every mark/.append style={fill=.!80!violettblau,solid},mark=otimes*\\%
lila, mark=|\\%
lila, dashed, mark=|\\%
gelb, every mark/.append style={fill=.!80!gelb},mark=pentagon*\\%
gelb, dashed, every mark/.append style={fill=.!80!gelb,solid},mark=pentagon*\\%
}

\IEEEoverridecommandlockouts

\usetikzlibrary{external}
\begin{document}

\title{Serial vs. Parallel Turbo-Autoencoders and Accelerated Training for Learned Channel Codes}

\author{
\IEEEauthorblockN{Jannis Clausius\IEEEauthorrefmark{1}, Sebastian D\"orner\IEEEauthorrefmark{1}, Sebastian Cammerer\IEEEauthorrefmark{2}\thanks{Work carried out while S. Cammerer was with the University of Stuttgart.} and Stephan ten Brink\IEEEauthorrefmark{1}\\}

\IEEEauthorblockA{\IEEEauthorrefmark{1}
Institute of Telecommunications, Pfaffenwaldring 47, University of  Stuttgart, 70569 Stuttgart, Germany
}
\IEEEauthorblockA{\IEEEauthorrefmark{2}NVIDIA, Fasanenstra{\ss}e 81, 10623 Berlin, Germany \\ \{clausius,doerner,tenbrink\}@inue.uni-stuttgart.de, scammerer@nvidia.com}

\thanks{This work has been supported in part by the DFG, Germany, under grant BR 3205/6-1 and in part by the Federal Ministry of Education and Research of the Federal Republic of Germany through the FunKI project under grant 16KIS1187.}
}

\maketitle

\begin{abstract}
Attracted by its scalability towards practical codeword lengths, we revisit the idea of Turbo-autoencoders for end-to-end learning of PHY-Layer communications. For this, we study the existing concepts of Turbo-autoencoders from the literature and compare the concept with state-of-the-art \emph{classical} coding schemes. We propose a new component-wise training algorithm based on the idea of Gaussian a priori distributions that reduces the overall training time by almost a magnitude. Further, we propose a new serial architecture inspired by \emph{classical} serially concatenated Turbo code structures and show that a carefully optimized interface between the two component autoencoders is required.
To the best of our knowledge, these serial Turbo autoencoder structures are the best known neural network based learned sequences that can be trained from scratch without any required expert knowledge in the domain of channel codes.

\end{abstract}

\acresetall

\section{Introduction}

Deep learning for PHY-Layer communications (and beyond) has become an attractive and active area of research ranging from channel estimation \cite{shlezinger2020viterbinet} and decoding \cite{nachmani2016learning}, \ac{MIMO} detection \cite{samuel2017deep} to fully learned transceiver designs \cite{8054694}.
In particular, the end-to-end learning idea \cite{o2016learning} promises to jointly learn the complete data-link between transmitter and receiver including all required signal processing such as detection, equalization and decoding \cite{Doerner2018}. As such, autoencoders promise a paradigm change in how we design future communication systems.
However, the current autoencoder neural network architectures do not scale well and are typically limited by an exponential training complexity (cf. \cite{gruber2017}). 
Thus, we believe structures that scale well towards longer message sequences are of great importance and a deeper analysis of the underlying effects, including the training, must be conducted.

In the literature, one can essentially distinguish between approaches that learn from scratch \cite{gruber2017,jiang2019turbo,kim2020deepcode,jiang2020learn} and the idea of embedding domain expert knowledge, i.e., fixing a \emph{macroscopic} structure and only learn specific parts of an algorithm \cite{nachmani2016learning,shlezinger2020viterbinet}. For example, in \cite{cammerer2020trainable} we have tackled the scaling problem by providing an explicit graph structure stemming from an outer \ac{LDPC} code that can be interpreted as explicit neural network following the idea of \emph{deep unfolding} \cite{nachmani2016learning}. Although the concept works well, we have to rely on the success of \emph{classical} code construction techniques that are not directly driven by real-world data.  
For this, the authors of \cite{jiang2019turbo} have proposed a Turbo code inspired neural network structure that even outperforms classical algorithms in some scenarios. It turns out that the Turbo-autoencoder structure can be easily extended to scenarios where no suitable solution is known or adaptivity w.r.t. different noise distributions is desired  \cite{jiang2019turbo} or to codes with feedback \cite{kim2020deepcode}. 
Other approaches \cite{shlezinger2020viterbinet} leverage the Viterbi algorithm by deep learning techniques.

In this work, we follow the same concept as in \cite{jiang2019turbo} and propose a new serially concatenated Turbo-autoencoder \ac{NN} structure. We analyze the differences in terms of training capabilities, achievable performance and scalability.
As the training of Turbo autoencoders still takes comparatively long -- although it does not scale exponentially -- we propose a pre-training technique that effectively lowers the training time by almost a magnitude. The concept is based on the idea of Gaussian distributed priors.
We want to emphasize that our main objective of this work is not necessarily to outperform existing coding schemes, but to learn about \ac{NN} structures that enable scalability towards arbitrary block lengths. As the field of channel coding provides a well-defined sandbox with the possibility of clear benchmarks and performance bounds, we think it is a fruitful scenario to learn about network architectures and their scalability.
Further, it is one step towards interpreting and understanding neural networks for communications in terms of their iterative information flow instead of plain universal function approximators.

\section{Parallel Turbo-Autoencoder}

In the following, we will shortly revisit the TurboAE \cite{jiang2019turbo} idea and its \ac{NN} structure based on the concept of parallel Turbo codes \cite{berrou1993near}. Please note that this simplifies the later description of the proposed serial structure but it is also required to clarify the notation for the proposed new training algorithm. %

\subsection{System Model}
A simple communication system can be modeled by an encoder%
, a channel %
and a decoder. %
The encoder %
maps a bit vector $\uv \in \mathbb{F}_{2}^{k}$ of length $k$ onto a real baseband symbol vector $\xv \in \RR^{n}$ of length $n$.\footnote{As often done in channel coding, we define the channel as real-valued. However, extensions towards complex-valued symbols are straightforward \cite{Doerner2018}.} The length defines the communication/coding rate of the system as $r=\frac{k}{n}$.
The channel  adds noise to the symbols.\footnote{In this work, we neglect other channel effects such as distortion or multi-path propagation.} %
Throughout this paper we use an \ac{AWGN} channel $\yv=\xv+\zv$ where the noise $\zv$ follows the distribution $\zv \sim \Nc\LB \vec{0}, \sigma^2 \vec{I}\RB$ with noise power $N_0 = 2\sigma^2$. The bit-wise \ac{SNR} of the system is defined as $\frac{E_b}{N_0}= \frac{1}{2r\sigma^2}$.
The decoder's task is to provide an estimate $\hat{\uv}$ of the transmitted bits $\uv$ based on the observed symbols $\yv$.

The idea of applying deep learning and a joint transceiver design \cite{8054694} is to replace the encoder and the decoder \ac{NN}s and train the resulting autoencoder in an end-to-end manner. However, one of the key problems is that the training complexity of the underlying classification problem scales with $2^{k}$. As a result, architectural restrictions are necessary, e.g., a Turbo structure combined with a \ac{CNN} to overcome this so-called \emph{curse of dimensionality}.

\begin{figure}[]
	\centering
	\resizebox{!}{!}{\input{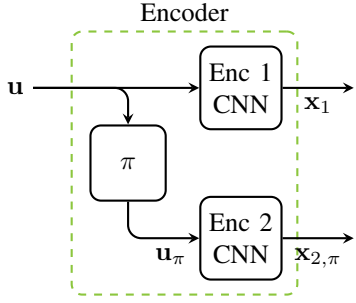}}
	\caption{Parallel concatenated encoder structure for $r=\frac{1}{2}$.}

	\label{fig:parallel_encoder}
	 \vspace*{-0.4cm}
\end{figure}

\begin{figure}[]
	\centering
	
	\resizebox{0.4\textwidth}{!}{\input{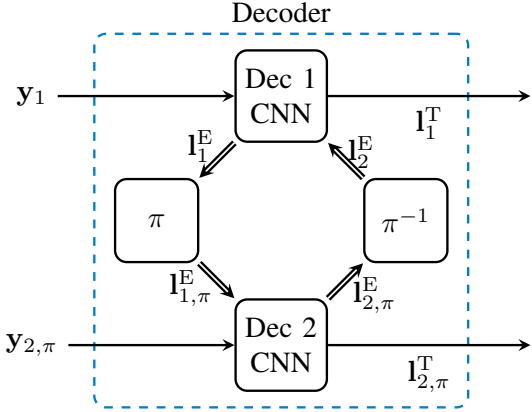}}
	\caption{Parallel Turbo decoder structure for $r=\frac{1}{2}$.}

	\label{fig:parallel_decoder}
	 \vspace*{-0.4cm}
\end{figure}

\subsection{Parallel Neural Network Structures}
We will briefly introduce the TurboAE concept \cite{jiang2019turbo}, which is the inspiration and basis of this work. The TurboAE uses \acp{CNN} as encoder and decoder in a parallel Turbo code structure. For more information on the design of the \acp{NN} and how to train the structure we refer to \cite{jiang2019turbo}. 
Parallel Turbo codes consist of the parallel concatenation of multiple encoders and an iterative decoding procedure with multiple decoding components.
In this paper, we  will focus on Turbo code structures with rate $r=\frac{1}{2}$ as shown in Fig.~\ref{fig:parallel_encoder} and Fig.~\ref{fig:parallel_decoder}. The extension to lower rates is straightforward as shown in \cite{jiang2019turbo}.
For the parallel concatenation, one of the encoders sees the input bits $\uv$ and the other sees a pseudo-random interleaved version $\uv_\pi =\pi\LB\uv\RB$. Throughout the paper the subscript $\pi$ denotes that a vector is interleaved. 
A pseudo-random interleaver $\pi\LB\cdot\RB$ maps the $i$th entry of $\uv$ to the $j$th entry of $\uv_\pi$ with $i\neq j ~\forall~ i,j=1,...,k$. Further, a deinterleaver $\pi^{-1}\LB\cdot\RB$ is the  inverse mapping of an interleaver.

The decoder consists of two decoders that see different, noisy symbol vectors $\yv_1 = \xv_1+\zv_1$ and $\yv_{2,\pi} = \xv_{2,\pi}+\zv_2 $ as input and the respective a priori information of the other decoder. The a priori information of one decoder is the extrinsic information of the other decoder from the previous iteration, which is calculated by $\lv^{\mathrm{E}} = \lv^{\mathrm{T}} - \lv^{\mathrm{A}}$,
where $\lv^{\mathrm{E}}$ is the extrinsic information, $\lv^{\mathrm{T}}$ is the total information and $\lv^{\mathrm{A}}$ is the a priori information. Thus, the  decoder \acp{NN} have inherent residual connections. The interleavers ensure that the order of the entries of $\lv^{\mathrm{E}}$ aligns with the respective channel observations.
The iterative decoding process is initialized with no a priori information $\lv^{\mathrm{E}} = \Vec{0}$. After a sufficient amount of iterations the bit estimates are $\hat{\uv}=\mathrm{H}(\lv^{\mathrm{T}})$, where $\mathrm{H}(\cdot)$ is the element-wise unit step function, i.e., the output is hard-decided.
Note, the double arrows in Fig.~\ref{fig:parallel_decoder} (and all other figures) illustrate that  $\lv^{\mathrm{E}}$ is a matrix with dimensions $k\times F$. For $F=1$ every entry can be interpreted as soft bit estimate, which is similar to the concept of \ac{LLR}s that are defined as
\begin{equation}
    \forall i \in \LP1,...,k\RP, \mathrm{LLR}(i):=\log\frac{p\LB \hat{u}_i=1|\yv \RB}{p\LB \hat{u}_i=0|\yv \RB}
\end{equation}
This relationship enables the generation of \emph{artificial} a priori information to achieve a faster training for the parallel TurboAE as shown in Sec. IV.

\begin{figure}[t]
	\centering
	
	\resizebox{0.5\textwidth}{!}{\input{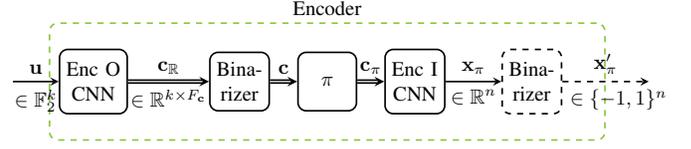}}
	\caption{Serial concatenated encoder structure with \emph{binarization} using \ac{STE}. Note, the second binarizer is optional.}

	\label{fig:serial_encoder}
	 \vspace*{-0.4cm}
\end{figure}

\begin{figure}[t]
	\centering
	
	\resizebox{0.4\textwidth}{!}{\input{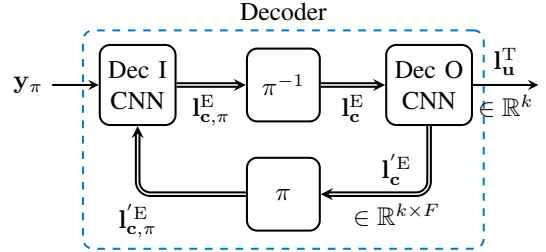}}
	\caption{Serial Turbo decoder}

	\label{fig:serial_decoder}
	 \vspace*{-0.4cm}
\end{figure}

\subsection{Modulation}
In contrast to a classical modulation scheme (e.g. \ac{BPSK}), every symbol $x_i$ in an \ac{NN}-based modulator
 might depend on the complete bit sequence, i.e., $\uv$, and not only on a small subset. The only restrictions for a neural modulator is the energy constraint $E_s=r\cdot E_b = \Expect{\xv}{\xv^2}=1$, thus the modulation order could be (hypothetically) up to $2^k$, which might result in modulation gains compared to a lower order modulation. For a distinction between coding gain and modulation gain, a neural modulator can be forced to a \ac{BPSK} modulation through binarization of the symbols $\xv$. As a result of the binarization, the autoencoder is no longer differentiable and a surrogate model must be introduced. In \cite{jiang2019turbo} a saturated \ac{STE} \cite{courbariaux2016binarized} was used which implements $\xv'=\mathrm{sign}(\xv)$ in the forward pass and implements the gradient:
\begin{equation}
\frac{\partial \xv'}{\partial \xv} = \begin{cases}
    \vec{1} &\text{for $|\xv|<1$}\\
    \vec{0} &\text{else.}
    \end{cases}
\end{equation}

We want to emphasize that quantization/binarization of \acp{NN} is still an active field of research and many different approaches exists (e.g., \emph{stochastic binarization} as in \cite{courbariaux2016binarized}). We follow the approach from \cite{jiang2019turbo} and leave it open for future research to further improve the binarization technique.

\section{Serial Turbo Autoencoder}

In this section we propose and analyze a new architecture based on serial Turbo codes as show in Fig.~\ref{fig:serial_encoder} and Fig.~\ref{fig:serial_decoder}. Similar to the approach of the TurboAE for parallel Turbo codes, we replace the encoders and decoders of a serial Turbo code with \acp{NN}. It turns out, that a carefully designed interface between the two component encoders/decoders is required.

\subsection{From Parallel to Serial Architectures}
The first encoder, which is called outer encoder, encodes the bits $\uv$ into a coded sequence $\cv_{\RR}$, where the subscript indicates that every entry is a real value. After that, a binarizer (realized as \ac{STE}) follows.
While this binarizer is no inherent part of the serial Turbo structure and seems like an unncessary restriction, it leads to a faster convergence during training and results in a  lower final \ac{BER}. Subsequently, a pseudo-random interleaver interleaves $\cv$ which is then used as input to the so called inner encoder. The inner encoder maps the coded sequence $\cv_\pi$ to a symbol sequence $\xv_\pi$. Additionally, $\xv_\pi$ could be binarized as well if a binary code is required.

The serial iterative decoding differs in two ways from the parallel iterative decoding. (1)  Only the inner decoder  sees the observed sequence $\yv$ and the outer decoder  sees no channel observations. (2) The outer decoder has two outputs. First, it outputs the extrinsic information of the coded sequence $\lv_{\cv}^{\mathrm{E}}$. Second, it outputs the total information of the bit sequence $\lv_{\uv}^{\mathrm{T}}$.
A consequence of the new structure is that the  design parameter $F_{\cv}$ which refers to the number of features of $\cv_\RR\in\RR^{k\times F_{\cv}}$ (illustrated by double arrows in Fig.~\ref{fig:serial_encoder}). Thus, the length of the interleaver is no longer forced to equal $k$, but can be up to $k\cdot F_{\cv}$. However, while we found that $F_{\cv}$ impacts the performance of the system, the interleaver length shows no impact. Empirically, a matching feature size $F_{\cv}= F$ between encoder and decoder works well.
We want to emphasize that the exact choice of $F_{\cv}$ has only an impact on the internal data-flow during encoding/decoding, however, it has no impact on $\xv$ nor on the communication rate $r$.

In addition to the empirical observation that the binarization of $\cv$ improves the performance, we want to provide an intuition of why this is the case. 
Without the binarization, the coded sequence is $\cv\in \RR^{k\times F_{\cv}}$. As a result, we observed that the estimation of the decoder messages  $\lv_{\cv}$ can be seen as a regression task, which means $\lv_{\cv}$ converges to a scaled version of $\cv$. %
In contrast, a binary  $\cv$ allows $\lv_{\cv}$ to behave like binary-soft-estimates of a classification task, which might be more meaningful for the internal decoding data-flow. 

\subsection{Results}
\begin{figure}[t]
	\centering
	\includegraphics{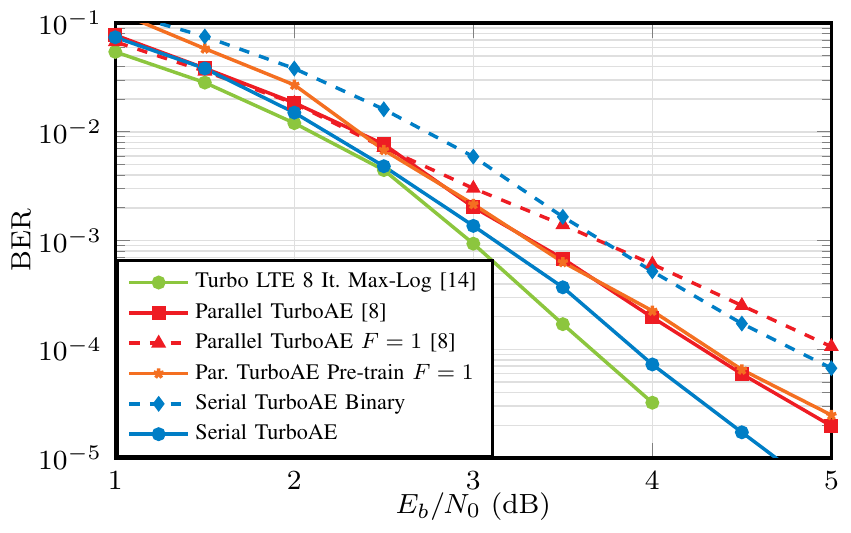}
	\caption{\ac{BER} achieved by TurboAE variants and the Turbo LTE code for $(k,n)=(64,128)$.}
	\label{fig:ber}
	\vspace*{-0.4cm}
\end{figure}

\begin{figure}[t]
	\centering
	\vspace*{-0.4cm}
	\includegraphics{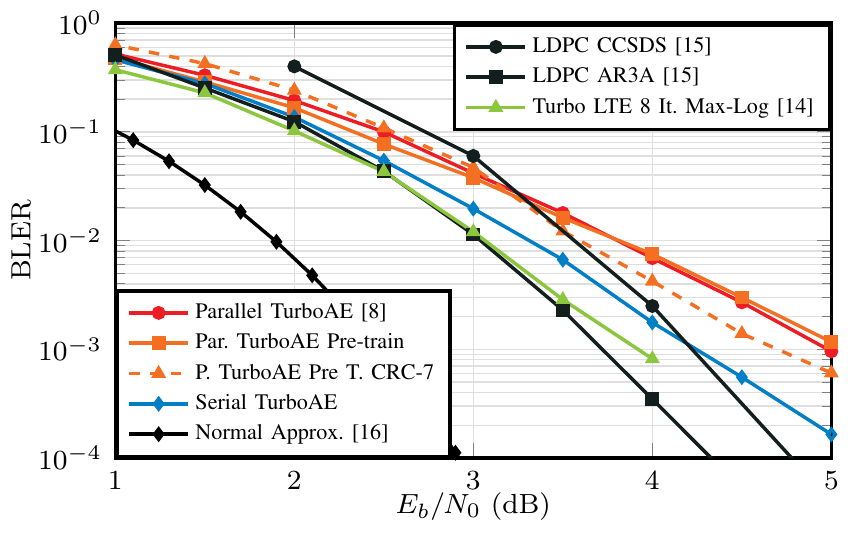}
	\caption{\ac{BLER} achieved by TurboAE variants and reference curves for $(k,n)=(64,128)$.}
	\label{fig:bler}
	\vspace*{-0.4cm}
\end{figure}

\begin{figure}[t]
	\centering
	\includegraphics{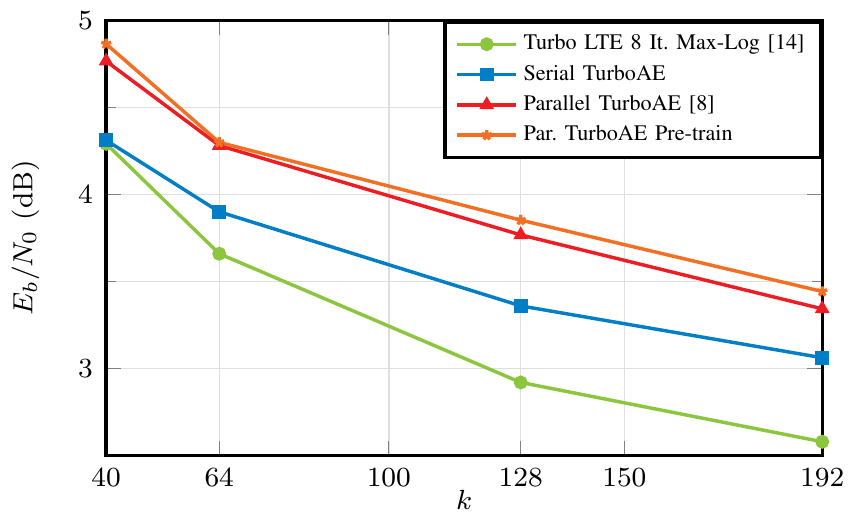}
	\caption{\ac{SNR} at a target \ac{BER} of $10^{-4}$ for several block lengths and $r=\frac{1}{2}$.}
	\label{fig:ksweep}
	\vspace*{-0.4cm}
\end{figure}

Before we continue with the results, we want to emphasize again, that this work is not about outperforming state-of-the-art methods and rather about introducing scalable \ac{NN}-based alternatives.
The \ac{BER} results are shown in Fig.~\ref{fig:ber} for $k=64$, $r=\frac{1}{2}$ and $F_{\cv}= F=10$. The serial structure shows significant gains compared to the parallel one, but still shows a gap towards the LTE Turbo code \cite{3gpp2007standard}. Furthermore, the figure includes the result of the training method for the parallel TurboAE that we propose in Sec. IV. Note, the method requires the sub-optimal choice of $F=1$, which leads to significant loss for the parallel TurboAE without Gaussian pre-training. Finally, the serial autoencoder with binary modulation is shown. Interestingly, the binary modulation at $k=64$ and $r=\frac{1}{2}$ leads to a higher loss (compared to the modulation which is only energy constrained) than for $k=100$ and $r=\frac{1}{3}$ as shown in \cite{jiang2019turbo}.
The resulting \acp{BLER} are shown in Fig.~\ref{fig:bler}, where they are compared with several well-known state-of-the-art codes from \cite{liva2016code} and the normal approximation \cite{polyanskiy2010convergence}. Note that the autoencoders are optimized for \ac{BER}. However, the \ac{BLER} of the serial structure shows a competitive performance.
The difference in terms of \ac{BLER} of the serial and parallel autoencoder is greater than the difference in terms of \ac{BER}. Our observation is that a significant amount of the block errors of the pre-trained parallel TurboAE origin from a single erroneous bit (at random codeword positions). To further improve the reliability, we embed a \ac{CRC}-7 code\footnote{The coding rate with the additional CRC-7 code is $r=0.445$ and is considered in the graph as an \ac{SNR} shift.} with additional bit-flip decoding in the structure  and show the results in Fig.~\ref{fig:bler}.
Finally, in Fig.~\ref{fig:ksweep} the required \ac{SNR} for a target \ac{BER} of $10^{-4}$ is shown for the block lengths $k=40,64,128,192$. For $k>64$ all autoencoders are finetuned with their respective $k$ on a trained model for $k=64$. In the case of $k=128$, the finetuning training from $k=64$ to $k=128$ leads to better results than a model for $k=128$ trained from scratch. The serial TurboAE shows the same performance as the LTE Turbo code for $k=40$, but the LTE Turbo code shows an increased coding gain for increasing $k$.

\begin{figure}[t]
	\centering
	\includegraphics{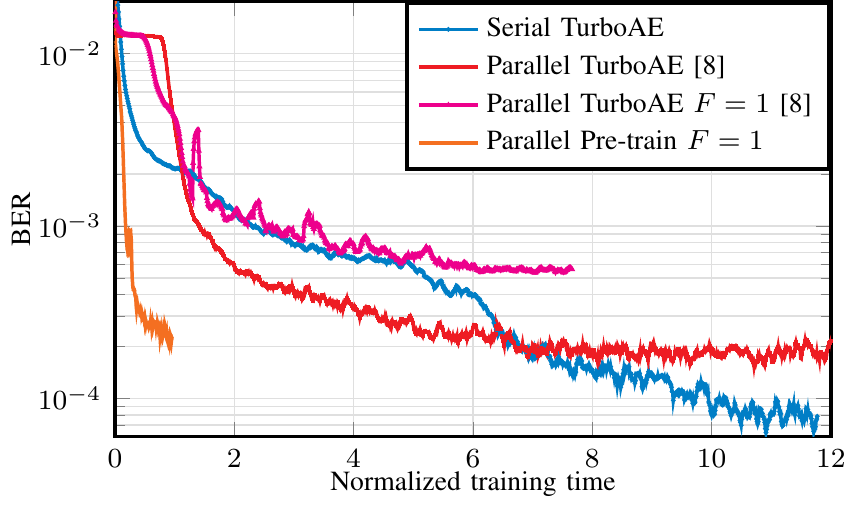}
	\caption{Moving average ($N=10$) of the \ac{BER} evaluation at an \ac{SNR} of $\frac{E_b}{N_0}=4\mathrm{dB}$ of several TurboAE variants for $(k,n) =(64,128)$. The training time is normalized on the duration of the pre-training for the parallel TurboAE ($1.85\mathrm{h}$ on three GTX 980).}
	\label{fig:loss}
	\vspace*{-0.4cm}
\end{figure}

\begin{figure}[t]
	\centering
	
	\resizebox{!}{!}{\input{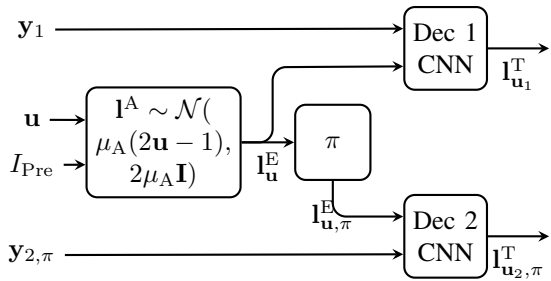}}
	\caption{Decoder for component-wise training via Gaussian priors. It is \emph{only} used for training.}

	\label{fig:parallel_decoder_pre_train}
	 \vspace*{-0.4cm}
\end{figure}

\section{Improved Training Strategy and Interpretation}

The same training algorithm that was proposed for the parallel TurboAE \cite{jiang2019turbo} can be applied to the serial TurboAE as well. Tab.~\ref{tab:hyperparameter} summarizes the parameters.
We now analyze the training complexity and provide a pre-training method to keep the training feasible based on the idea of Gaussian distributed a priori information.
    \begin{table}
	\caption{Hyperparameters for the training algorithm from \cite{jiang2019turbo}.}\label{tab:hyperparameter}
	\centering
	\begin{tabular}{l|l}
	
		Parameter& Value \\ 
		\hline
		$(k,n)$ & $(64,128)$ \\ 
		$F_{\cv} = F$ & $10$   \\
		$T_{\mathrm{enc}}$  & $100$  \\
		$T_{\mathrm{dec}}$ &  $500$  \\
		Decoder Iterations & $6$

	\end{tabular}
	\begin{tabular}{l|l}
	
		Parameter& Value \\ 
		\hline
        Optimizer & ADAM  \\
		Batchsize & $500-2000$  \\
		Learning rate & $10^{-4}-10^{-6}$  \\
		Encoder \ac{SNR} & $4.0\mathrm{dB}$  \\
		Decoder \ac{SNR} & $0.5-4.0\mathrm{dB}$
		
	\end{tabular}
	\end{table}

We observed that the training time scales roughly linear with the block length $k$. As a result,  the training is not feasible for $k\gtrsim128$. A speedup can be achieved by training a model at $k=64$ and subsequently increase $k$ to the desired block length.  
As can be seen in Fig.~\ref{fig:loss} another significant speedup is achieved by the Gaussian pre-training.

\subsection{Gaussian Pre-training}\label{sec:GA_training}

We provide a new training algorithm for the parallel TurboAE that is -- compared to the algorithm from \cite{jiang2019turbo} -- roughly $6.5$ times faster at the price of a slightly worse \ac{BER} performance. The structure of the decoder during training (only during training!) is show in Fig.~\ref{fig:parallel_decoder_pre_train}, while the final decoder structure during decoding inference remains unchanged. The idea is to train encoder/decoder-1 separately from encoder/decoder-2 by decoupling of the two components and approximating the a priori information by Gaussian distributed \acp{LLR} \cite{tenbrink2001concatcodes}. 
As a result, we can train the encoder and the decoder with a single decoder iteration, which significantly decreases the time for training.
The first consequence of the Gaussian pre-train is that the feature size of the messages is set to $F=1$.\footnote{Please note that the approach may work for other values than $F=1$ but the underlying distribution must be known/approximated. However, we leave it open for future research to find such a priori distributions for $F>1$.}
The second consequence is that the decoder shares the weights over iterations during normal operation. Before, we could have different weights specialized for their respective iteration. 
Therefore, the decoder is not finetuned for a certain amount of iterations. That means, more iterations lead to a lower \ac{BER}; we used $96$ iterations during inference. In contrast, more than $6$ decoder iteration in combination with the training from \cite{jiang2019turbo} increases the \ac{BER}.

The generation of a priori \acp{LLR} $\lv^{\mathrm{A}}$ of the bits $\uv$ is based on the assumption that they are Gaussian distributed and do not correlate with the channel observations $\yv_i ,~i=1,2$. If these assumptions hold, the a priori \acp{LLR} of the bit sequence $\uv$ are $ \lv^{\mathrm{A}} \sim \Nc \LB  \mu_{\mathrm{A}}(2\uv-1),2\mu_{\mathrm{A}} \vec{I} \RB $ distributed with
\begin{align}
    \mu_{\mathrm{A}} &= \mathrm{J}^{-1}(I_{\mathrm{Pre}}) \approx \LB - \frac{1}{H_1}\log_2\LB 1-I_{\mathrm{Pre}}^{\frac{1}{H_3}} \RB \RB^{\frac{1}{2H_2}} 
\end{align}
where $\mathrm{J}^{-1}$ is the inverse J-function and the coefficients $H_1 =0.3073$, $H_2 =0.8935$ and $H_3 =1.1064$ \cite{brannstrom2005convergence}.

\subsection{Hyperparameters}
The only change to the training hyperparameters in Tab.~\ref{tab:hyperparameter} is an increase of the decoder updates per epoch to $T_{dec}=1000$. 
Additionally, a new  parameter is the choice of $I_{\mathrm{Pre}}$ and its different values during encoder and decoder training. The choice during encoder training needs to be carefully considered with regards to the desired result. In contrast,  for decoder training the following choice seems robust. Since the decoder needs to be able to function with any amount of prior information it can be uniformly distributed  $I_{\mathrm{Pre}}\sim \mathcal{U}(0,1)$.
However, more attention is required during encoder training. On the one hand, $I_{\mathrm{Pre}}$ must be small enough so that the encoder learns a mapping that enables the iterative decoding in the first iteration without prior knowledge.
On the other hand, $I_{\mathrm{Pre}}$ must be large enough such that the decoder learns how to deal with a priori information.
Our observation is that a higher $I_{\mathrm{Pre}}$ leads to a steeper \ac{BER} curve. Particularly at the beginning of the training process, a high $I_{\mathrm{Pre}}$ speeds up the convergence, but $I_{\mathrm{Pre}}$ might need to be adjusted later.
In the case of $k=64$, we started with $I_{\mathrm{Pre}}\sim \mathcal{U}(0.8,1)$ and finetuned with  with $I_{\mathrm{Pre}}\sim \mathcal{U}(0.5,1)$.

\section{Conclusion and Outlook}
Turbo-autoencoders have the potential to scale the end-to-end learning idea towards practical block lengths.
Although, we currently only compete -- but do not necessarily outperform -- classical algorithms, we are attracted by the vision of learning PHY-layer communications from scratch.
We have presented a new Turbo-autoencoder structure based on the concept of serial concatenated Turbo codes. 
Further, we have shown that pre-training with Gaussian priors greatly accelerates the training without significantly losing performance.
We believe that dedicated \ac{NN} structures -- such as the TurboAE -- tailored to problems in communications are an important step towards practical applications of deep learning for communications and, as such, the TurboAE may become an important milestone on the path towards future \emph{learning-defined} systems.

\section*{Acknowledgments}
We would like to thank Marvin Geiselhart for providing the Turbo code reference simulations.

\bibliographystyle{IEEEtran}
\bibliography{IEEEabrv,references}

\end{document}